\documentclass[12pt,preprint]{aastex}
\makeatletter
  
  \@addtoreset{equation}{section}
\makeatother

\slugcomment{The Astrophysical Journal, Submitted 2002 May}

\shortauthors{Itoh et al.}
\shorttitle{SCREENING CORRECTIONS TO THE ELECTRON CAPTURE RATES}

\begin{document}

\title{SCREENING CORRECTIONS TO THE ELECTRON CAPTURE RATES
 IN DENSE STARS BY THE RELATIVISTICALLY DEGENERATE ELECTRON LIQUID}

\author{\sc Naoki Itoh, Nami Tomizawa, Masaya Tamamura,
 and Shinya Wanajo}
\affil{Department of Physics, Sophia University,
       7-1 Kioi-cho, Chiyoda-ku, Tokyo, 102-8554, Japan;\\
       n\_itoh@sophia.ac.jp, tomiza-n@sophia.ac.jp, 
       m-tamamu@sophia.ac.jp, wanajo@sophia.ac.jp}

\and

\author{\sc Satoshi Nozawa}
\affil{Josai Junior College for Women, 1-1 Keyakidai, Sakado-shi,
       Saitama, 350-0295, Japan; snozawa@josai.ac.jp}

\bigskip
\affil{The Astrophysical Journal, Submitted 2002 May}

\begin{abstract}

We calculate the screening corrections to the electron capture rates in
dense stars by the relativistically degenerate electron liquid.  In order
to calculate the screening corrections we adopt the linear response theory
which is widely used in the field of solid state physics and liquid metal
physics.  In particular, we use the longitudinal dielectric function for
the relativistically degenerate electron liquid derived by Jancovici.
We calculate the screening potential at the position of the nucleus.
By using this screening potential one can calculate the screening
corrections to the electron capture rates.  We will present accurate
analytic fitting formulae which summarize our numerical results.  These
fitting formulae will facilitate the application of the present results.
The screening corrections to the electron capture rates are typically a few
percent.

\end{abstract}


\section{INTRODUCTION}

Since the pioneering works of Fuller, Fowler, \& Newman (1980, 1982a,
1982b, 1985), calculations of stellar weak-interaction rates have entered
an era of precision science.  More recently an authoritative work of
Langanke \& Martinez-Pinedo (2000) on this subject appeared.

Since the presupernova stellar evolution and stellar nucleosynthesis
critically depend on the details of the stellar weak-interaction
rates (e.g., Wanajo et al. 2002), it is extremely important to
calculate accurately the screening corrections to the electron
capture rates in dense stars by the relativistically degenerate
electron liquid.

This problem has been already addressed by many authors (Couch \& Loumos
1974; Takahashi, El Eid, \& Hillebrandt 1978; Gutierrez et al. 1996;
Luo \& Peng 1996; Bravo \& Garcia-Senz 1999).  The plasma effects on the
chemical potential of the nucleus and hence on the threshold energy for
the electron capture, in particular, have been discussed by Couch \&
Loumos (1974), Gutierrez et al. (1996), as well as by Bravo \&
Garcia-Senz (1999).

In this paper we will address ourselves to the calculation of the
effective potential energy felt by the relativistically degenerate
electron.  We will use the linear response theory in order to calculate
the screening potential caused by the relativistically degenerate
electron liquid.

The present paper is organized as follows.  In \S~2 we will calculate
the effective potential energy felt by the electron using the
longitudinal dielectric function of the relativistically degenerate
electron liquid derived by Jancovici (1962).  We will thereby calculate
the screening potential which will be used for the calculation of the
screening corrections to the electron capture rates.  In \S~3 we will
summarize the numerical results in the form of analytic fitting
formulae which will facilitate the application of the present results.
We will give concluding remarks in \S~4.

\section{SCREENING OF THE COULOMB POTENTIAL BY THE RELATIVISTICALLY
DEGENERATE ELECTRON LIQUID}

The Coulomb corrections to the beta decay rates have been discussed in
many standard textbooks on beta decay (e.g., Morita 1973).  Some authors
(Takahashi, El Eid, \& Hillebrandt 1978; Fuller, Fowler, \& Newman 1980)
have discussed possible importance of the screening effects on the
electron capture rates at extremely high densities.  Therefore, it is
extremely important to calculate accurately the screening corrections
to the electron capture rates by the relativistically degenerate
electron liquid.

We consider the density-temperature regime in which the electrons
are strongly degenerate.  This condition is expressed as (Itoh et al.
1983)
\begin{eqnarray}
T \ll T_{F} & = & 5.930 \times 10^{9} \, \left[ \left\{ 1 + 1.018 
 (Z/A)^{2/3} \rho_{6}^{2/3} \right\}^{1/2} - 1 \right] \,  [{\rm K}] \, ,
\end{eqnarray}
where $T_{F}$ is the electron Fermi temperature, $\rho_{6}$ is the
mass density in units of $10^{6}$ g cm$^{-3}$, and $Z$ and $A$ are the
atomic number and mass number of the nucleus considered, respectively.
We assume complete pressure ionization.  Therefore, no bound electrons
will exist around the nucleus.  This assumption is valid when the
following condition is satisfied (Itoh et al. 1996):
\begin{eqnarray}
\frac{Z}{A} \, \rho & \geq & 0.378 \, Z^{3} \, \, [ {\rm g \, cm}^{-3} ]
    \, ,
\end{eqnarray}
where $\rho$ is the mass density.

The potential energy of the electron due to the pure Coulomb potential
$-Ze^{2}/r$ is changed to $V(r)$ because of the screening by the
relativistically degenerate electron liquid.  In this paper we use the
static longitudinal dielectric function due to the relativistically
degenerate electron liquid calculated by Jancovici (1962) based on the
relativistic random-phase approximation.  Jancovici's static longitudinal
dielectric function is written in the form (Itoh et al. 1984)
\begin{eqnarray}
\epsilon(q, 0) & = & 1 \, + \, \left( \frac{2}{3 \pi^{2}} \right)^{2/3} 
   \frac{r_{s}}{q^{2}} \, \left[ \, \frac{2}{3} (1 + x^{2})^{1/2} \,
    - \, \frac{2 q^{2}x}{3} {\rm sinh}^{-1} x \,  \right.  \nonumber \\
  &  & \, \, + \, (1 + x^{2})^{1/2} \, \frac{x^{2}+1-3q^{2}x^{2}}{6 
       q x^{2}} \, {\rm ln} \left| \frac{1+q}{1-q} \right|  \nonumber \\
  &  & \, \, \left. + \, \frac{2 q^{2}x^{2}-1}{6qx^{2}} 
        (1 + q^{2}x^{2})^{1/2} 
       \, {\rm ln} \left| \frac{q(1 + x^{2})^{1/2}+(1 + q^{2}x^{2})^{1/2}}
          {q(1 + x^{2})^{1/2}-(1 + q^{2}x^{2})^{1/2}} \right|
           \, \right] , \\
 q & = & \frac{k}{2 k_{F}} \, ,  \\
 x & = & \frac{ \hbar k_{F}}{m_{e}c} \, = \, \frac{1}{137.036} 
         \left( \frac{9 \pi}{4} \right)^{1/3} \, r_{s}^{-1} \, ,  \\
 r_{s} & = & \frac{ \, a_{e} \, }{ \displaystyle{ \frac{ \hbar^{2}}
             {m_{e}e^{2}} }} = 1.388 \times 10^{-2} 
             \left( \frac{A}{Z \rho_{6}} \right)^{1/3} \, ,   \\
  &  &  \frac{4}{3} \pi a_{e}^{3} \, n_{e} \, = \, 1  \, ,
\end{eqnarray}
where $n_{e}$ is the electron number density.  The electron Fermi wavenumber
is expressed as
\begin{eqnarray}
k_{F} & = & 2.613 \times 10^{10} \, \left( \frac{Z}{A} \, \rho_{6} 
            \right)^{1/3} \, \, [{\rm cm}^{-1}] \, .
\end{eqnarray}
The ion-sphere radius $a_{i}$ is written as (Itoh \& Kohyama 1993)
\begin{eqnarray}
\frac{4}{3} \pi a_{i}^{3} \, n_{i} & = & 1  \, ,  \\
a_{i} & = & 0.7346 \times 10^{-10} \left( \frac{ \rho_{6}}{A} 
            \right)^{-1/3} \, \, [{\rm cm}] \, ,
\end{eqnarray}
where $n_{i}$ is the ion number density.  We also have the relationship
\begin{eqnarray}
k_{F} a_{i} & = & \left( \frac{ 9 \pi}{4} \right)^{1/3} \, Z^{1/3} \, .
\end{eqnarray}

The potential energy of the electron $V(r)$ which takes into account the
screening by the relativistially degenerate electron liquid is written as
\begin{eqnarray}
V(r) & = & - \frac{Z e^{2}}{2 \pi^{2}} \, \int \frac{ e^{i \vec{k}
 \cdot \vec{r}}}{k^{2} \, \epsilon(k, 0)} d^{3} k    \nonumber  \\
     & = & - \frac{Z e^{2}(2 k_{F})}{2 k_{F}r} \, \frac{2}{\pi} \, 
     \int_{0}^{\infty} \frac{ {\rm sin}
      [(2 k_{F}r)q]}{q \, \epsilon(q, 0)} \, d q \, .
\end{eqnarray}
Therefore the increment of the potential energy due to the screening by
the relativistically degenerate electron liquid is
\begin{eqnarray}
V_{s}(r) & \equiv & V(r) \, - \, \left( - \frac{Z e^{2}}{r} \right) 
               \nonumber    \\
         & = &  Z e^{2} (2k_{F}) \, \frac{1}{2k_{F}r} \left\{ 1  - 
                \frac{2}{\pi} \, \int_{0}^{\infty} \frac{ {\rm sin}
      [(2 k_{F}r)q]}{q \, \epsilon(q, 0)} \, d q  \right\}  \nonumber  \\
      & = & 7.525 \times 10^{-3} \, Z \left( \frac{Z}{A} \rho_{6} 
            \right)^{1/3} [{\rm MeV}] \frac{1}{2k_{F}r} 
            \left\{ 1  -  \frac{2}{\pi} \, \int_{0}^{\infty} 
            \frac{ {\rm sin} [(2 k_{F}r)q]}{q \, \epsilon(q, 0)} \, d q 
             \right\} .
\end{eqnarray}

Equation (2.12) is based on the linear response theory (e.g., Pines \& 
Nozi\`{e}res 1966).  Dharma-wardama \& Perrot (1982, 1990) have carried out
the density-functional study of hydrogen plasmas as well as the 
density-functional study of C, Si, and Ge metallic liquids and have
found that the detailed results of the density-functional calculations
of these systems are close to the results obtained by the linear
response theory.  They have also confirmed that the density-functional
theory as well as the linear response theory reproduce satisfactorily
the experimental results on the Ge metallic liquid, thereby proving
the excellence of these theories for this system.  This fact gives
great support to our use of the linear response theory in the present
problem.

Compared with the case of the pure Coulomb potential, the effective
electron energy at the position of the nucleus is increased by 
$V_{s}(0)$.  Hence the usual factor in the electron-capture rates
$p E F(Z, E)$ (Morita 1973; Fuller, Fowler, \& Newman 1980;
Langanke \& Martinez-Pinedo 2000) should be replaced by
\begin{eqnarray}
  \left\{ \left[ E-V_{s}(0) \right]^{2} - m_{e}^{2} \right\}^{1/2} 
  \left[ E-V_{s}(0) \right] \, F \left[Z, E-V_{s}(0) \right] \, .
\end{eqnarray}
At the same time the shift in the threshold energy for the electron
capture should be taken into account (Couch \& Loumos 1974; Gutierrez
et al. 1996; Bravo \& Garcia-Senz 1999).

In Figure~1 we show the function
\begin{eqnarray}
I & \equiv &  \frac{2}{\pi} \, \int_{0}^{\infty} \frac{ {\rm sin}
      [(2 k_{F}r)q]}{q \, \epsilon(q, 0)} \, d q
\end{eqnarray}
as a function of $R=2k_{F}r$ for various values of $r_{s}$.  In Figure~2
we show the function
\begin{eqnarray}
J & \equiv &  \frac{1}{2k_{F}r} \left\{ 1  -  \frac{2}{\pi} \, 
             \int_{0}^{\infty} \frac{ {\rm sin} [(2 k_{F}r)q]}{q \, 
             \epsilon(q, 0)} \, d q 
             \right\}
\end{eqnarray}
as a function of $R=2k_{F}r$ for various values of $r_{s}$.  In Figure~3
we enlarge Figure~2 for small values of $R=2k_{F}r$.  In Figure~4 we show
$J(R=0)$ as a function of $r_{s}$.  Equation (2.13)
can be rewritten as
\begin{eqnarray}
V_{s}(r) & = & 7.525 \times 10^{-3} \, Z \left( \frac{Z}{A} \rho_{6} 
            \right)^{1/3} \, J \, [{\rm MeV}] \, .
\end{eqnarray}

The nuclear radius $r_{nuc}$ can be expressed as (Morita 1973)
\begin{eqnarray}
r_{nuc} & = & 1.2 \, A^{1/3} \, [{\rm fm}] \, .
\end{eqnarray}
From equations (2.8) and (2.18) we have
\begin{eqnarray}
2 k_{F} r_{nuc} & = & 6.3 \times 10^{-3} \, Z^{1/3} \rho_{6}^{1/3} \, .
\end{eqnarray}
From Figure~3 we see that the function $J(R)$ is almost constant for
small values of $R$ which corresponds to $2k_{F}r_{nuc}$ given by
equation (2.19), when the mass density is not extremely high.  This
fact justifies our use of $V_{s}(0)$ for the screening potential
at the nuclear radius.  When the mass density becomes extremely high,
then one should use the value of the screening potential $V_{s}(r)$
by adopting the value of the function $J(R=2k_{F}r_{nuc})$.

The electron Fermi energy is given by
\begin{eqnarray}
E_{F} & = & 0.5110 \, \left[ \left\{ 1 + 1.018 (Z/A)^{2/3}
          \rho_{6}^{2/3} \right\}^{1/2} \, - \, 1 \right] \, [{\rm MeV}] \, .
\end{eqnarray}
When the electrons are extremely relativistic, we have
\begin{eqnarray}
\frac{V_{s}(r)}{E_{F}} & \simeq & 1.460 \times 10^{-2} \, Z \, J \, .
\end{eqnarray}
Therefore, we find that the screening potential at the origin $V_{s}(0)$
is typically a few percent of the electron Fermi energy.  Thus we conclude
that the screening corrections to the electron capture rates by the
relativistically degenerate electron liquid are not as great as
anticipated by Takahashi, El Eid, \& Hillebrandt (1978) and also by
Fuller, Fowler, \& Newman (1980).

In passing, it is interesting to compare our present detailed calculation
with the Fermi-Thomas model which is generally more crude than the
random-phase approximation that has been adopted in the present paper
(Pines and Nozi\`{e}res 1966).  In the Fermi-Thomas model the screening
potential at the origin is expressed as (Pines and Nozi\`{e}res 1966)
\begin{eqnarray}
\left[ V_{s}(0) \right]_{FT} & = & Z e^{2} \, k_{FT} \, ,
\end{eqnarray}
where $k_{FT}$ is the Fermi-Thomas screening wave number.  When the
electrons are extremely relativistic, the Fermi-Thomas screening wave
number is expressed as (Flowers \& Itoh 1976)
\begin{eqnarray}
\frac{k_{FT}}{2 k_{F}} & \simeq & \frac{1}{2} \left( \frac{4}{ \pi} 
\frac{1}{137.036} \right)^{1/2} \, = \, 0.0482.
\end{eqnarray}
By Comparing with Figures~2, 3, 4, we find that the prediction of the
Fermi-Thomas model is quite good (the accuracy is about 7\%) for the
present problem when the electrons are extremely relativistic.

\section{ANALYTIC FITTING FORMULAE}

In this section we will present accurate analytic fitting formulae
for $J(R)$ in order to facilitate the application of the numerical
results obtained in the present paper.  We have carried out the numerical
calculations of $J(R)$ for $0.00001 \leq r_{s} \leq 0.1$, $0 \leq R
\leq 50.0$.  We express the analytic fitting formula by
\begin{eqnarray}
J(r_{s}, R) & = & \sum_{i,j=0}^{10} a_{i j} \, s^{i} u^{j} \, ,  \\
 s & \equiv & \frac{1}{2} \left( {\rm log}_{10} r_{s} \, + \, 3 
              \right) \, ,  \\
 u & \equiv & \frac{1}{25.0} \left( R \, - \, 25.0 \right) \, .
\end{eqnarray}
The coefficients $a_{ij}$ are presented in Table~1.  The accuracy of the
fitting is generally better than 0.1\%.

The value of $J(r_{s}, R)$ for $R=0$ is of course obtained by the fitting
formulae (3.1), (3.2), (3.3).  For the sake of simplicity we will give
a separate fitting formula
\begin{eqnarray}
J(r_{s}, R=0) & = & \sum_{i=0}^{10} b_{i} \, s^{i} \, ,  \\
 s & \equiv & \frac{1}{2} \left( {\rm log}_{10} r_{s} \, + \, 3 
              \right) \, .
\end{eqnarray}
The coefficients $b_{i}$ are presented in Table~2.  The accuracy of the
fitting is generally better than 0.1\%.

\section{CONCLUDING REMARKS}

We have studied the screening corrections to the electron capture rates
by the relativistically degenerate electron liquid.  In particular,
we have calculated the screening potential caused by the relativistically
degenerate electron by using Jancovici's (1962) static longitudinal
dielectric function.  We have found that the screening potential is
typically a few percent of the electron Fermi energy.  Hence we
conclude that the screening corrections to the electron capture rates
at high densities are not as great as anticipated by Takahashi, El Eid, 
\& Hillebrandt (1978) and also by Fuller, Fowler, \& Newman (1980).

We have presented accurate analytic fitting formulae which will be
useful when one wishes to apply the present results to the calculations
of the screening corrections to the electron capture rates at high
densities.

\acknowledgments

We wish to thank K. Nomoto for drawing our attention to the present
important problem.  We are grateful to Y. Takada for clarifying the
validity of the linear response theory in the present problem.  We
also thank Y. Oyanagi for allowing us to use the least-squares 
fitting program SALS.  This work is financially supported in part 
by Grants-in-Aid of the Japanese Ministry of Education, Culture,
Sports, Science, and Technology under contracts 13640245, 13740129.

\clearpage

%__________________________________________________ Table 1
\begin{table*}
\footnotesize
\caption[]{Coefficients $a_{ij}$}
\begin{tabular}{crrrrrr} \hline

  & {\it j}=0 \, \, \, & {\it j}=1 \, \, \, & {\it j}=2  \, \, \, & 
    {\it j}=3  \, \, \, & {\it j}=4  \, \, \, & {\it j}=5  \, \, 
    \, \\ \hline

 {\it i}=0 &  2.80066E$-$2  &  $-$1.34650E$-$2  &  4.70157E$-$3  &  $-$1.62773E$-$3  &  3.57498E$-$4  &  2.77894E$-$3  \\

 {\it i}=1 &  2.91425E$-$4  &  $-$4.77037E$-$4  &  3.24480E$-$4  &  $-$1.84976E$-$4  &  1.26065E$-$4  &  3.14205E$-$4  \\

 {\it i}=2 &  3.71730E$-$4  &  $-$1.11205E$-$3  &  1.11414E$-$3  &  $-$7.49702E$-$4  &  4.08871E$-$4  &  7.01182E$-$4  \\

 {\it i}=3 &  $-$3.40043E$-$4  &  1.76471E$-$3  &  $-$7.27685E$-$4  &  $-$2.02979E$-$3  &  3.21763E$-$3  &  4.64229E$-$3  \\

 {\it i}=4 &  8.38363E$-$3  &  $-$3.40534E$-$3  &  $-$1.66683E$-$3  &  $-$4.20858E$-$3  &  1.00310E$-$2  &  2.03169E$-$2  \\

 {\it i}=5 &  2.98675E$-$2  &  $-$3.93466E$-$2  &  1.81463E$-$2  &  $-$6.64180E$-$3  &  9.00760E$-$3  &  2.90509E$-$2  \\

 {\it i}=6 &  1.44775E$-$2  &  $-$4.06027E$-$2  &  3.23311E$-$2  &  $-$8.65196E$-$3  &  $-$2.12560E$-$3  &  1.34492E$-$3  \\

 {\it i}=7 &  $-$3.96957E$-$2  &  4.02288E$-$2  &  $-$5.91508E$-$3  &  $-$8.13296E$-$3  &  $-$1.31089E$-$4  &  $-$2.02386E$-$2  \\

 {\it i}=8 &  $-$3.32815E$-$2  &  5.69571E$-$2  &  $-$3.05845E$-$2  &  $-$1.66852E$-$3  &  7.99359E$-$3  &  $-$1.02361E$-$3  \\

 {\it i}=9 &  1.51205E$-$2  &  $-$1.18443E$-$2  &  $-$2.51457E$-$3  &  4.25544E$-$3  &  2.57406E$-$3  &  1.07350E$-$2  \\

 {\it i}=10 &  1.52622E$-$2  &  $-$2.14839E$-$2  &  8.10578E$-$3  &  2.56649E$-$3  &  $-$1.54054E$-$3  &  3.10744E$-$3  \\ \hline

\end{tabular}

\smallskip

\begin{tabular}{crrrrr} \hline

  & {\it j}=6 \, \, \, & {\it j}=7 \, \, \, & {\it j}=8  \, \, \, & 
    {\it j}=9  \, \, \, & {\it j}=10   \, \, \, \\ \hline

 {\it i}=0 &  $-$1.84849E$-$3  &  $-$5.69873E$-$3  &  4.05787E$-$3  &  4.57584E$-$3  &  $-$3.62782E$-$3  \\

 {\it i}=1 &  $-$4.25228E$-$4  &  $-$5.70927E$-$4  &  6.77612E$-$4  &  3.74132E$-$4  &  $-$4.07058E$-$4  \\

 {\it i}=2 &  $-$7.64487E$-$4  &  $-$1.83728E$-$3  &  1.70005E$-$3  &  1.40476E$-$3  &  $-$1.30179E$-$3  \\

 {\it i}=3 &  $-$7.73692E$-$3  &  $-$1.08840E$-$2  &  1.41242E$-$2  &  6.32205E$-$3  &  $-$7.84898E$-$3  \\

 {\it i}=4 &  $-$3.47978E$-$2  &  $-$3.75841E$-$2  &  5.56369E$-$2  &  1.91593E$-$2  &  $-$2.75063E$-$2  \\

 {\it i}=5 &  $-$4.92450E$-$2  &  $-$4.46316E$-$2  &  7.19123E$-$2  &  2.38135E$-$2  &  $-$3.55769E$-$2  \\

 {\it i}=6 &  $-$2.51385E$-$3  &  3.37337E$-$3  &  $-$1.61833E$-$3  &  3.64315E$-$3  &  $-$2.90671E$-$3  \\

 {\it i}=7 &  3.68866E$-$2  &  2.53148E$-$2  &  $-$5.16877E$-$2  &  $-$8.55152E$-$3  &  2.08971E$-$2  \\

 {\it i}=8 &  5.80492E$-$3  &  $-$1.04420E$-$2  &  $-$3.04257E$-$5  &  3.95689E$-$3  &  $-$2.15731E$-$4  \\

 {\it i}=9 &  $-$2.02777E$-$2  &  $-$1.85917E$-$2  &  3.19260E$-$2  &  7.62300E$-$3  &  $-$1.40661E$-$2  \\

 {\it i}=10 &  $-$8.44178E$-$3  &  $-$2.77226E$-$3  &  1.11309E$-$2  &  1.36109E$-$3  &  $-$5.00155E$-$3  \\ \hline

\end{tabular}
\end{table*}

\clearpage

%__________________________________________________ Table 2
\begin{table}
\caption[]{Coefficients $b_{i}$}
\begin{tabular}{cr} \hline

       &   $b_{i}$  \, \, \, \, \,   \\  \hline
       {\it i}=0   &     4.50861E$-$2    \\
       {\it i}=1   &     1.13078E$-$3    \\
       {\it i}=2   &     3.12104E$-$3    \\
       {\it i}=3   &     8.64302E$-$4    \\
       {\it i}=4   &     1.57214E$-$2    \\
       {\it i}=5   &     8.16962E$-$2    \\
       {\it i}=6   &     7.84921E$-$2    \\
       {\it i}=7   &  $-$6.80863E$-$2    \\
       {\it i}=8   &  $-$9.79967E$-$2    \\
       {\it i}=9   &     2.04907E$-$2    \\
       {\it i}=10  &     3.66713E$-$2    \\   \hline

\end{tabular}
\end{table}

\clearpage
\begin{figure}
\epsscale{0.8}
\plotone{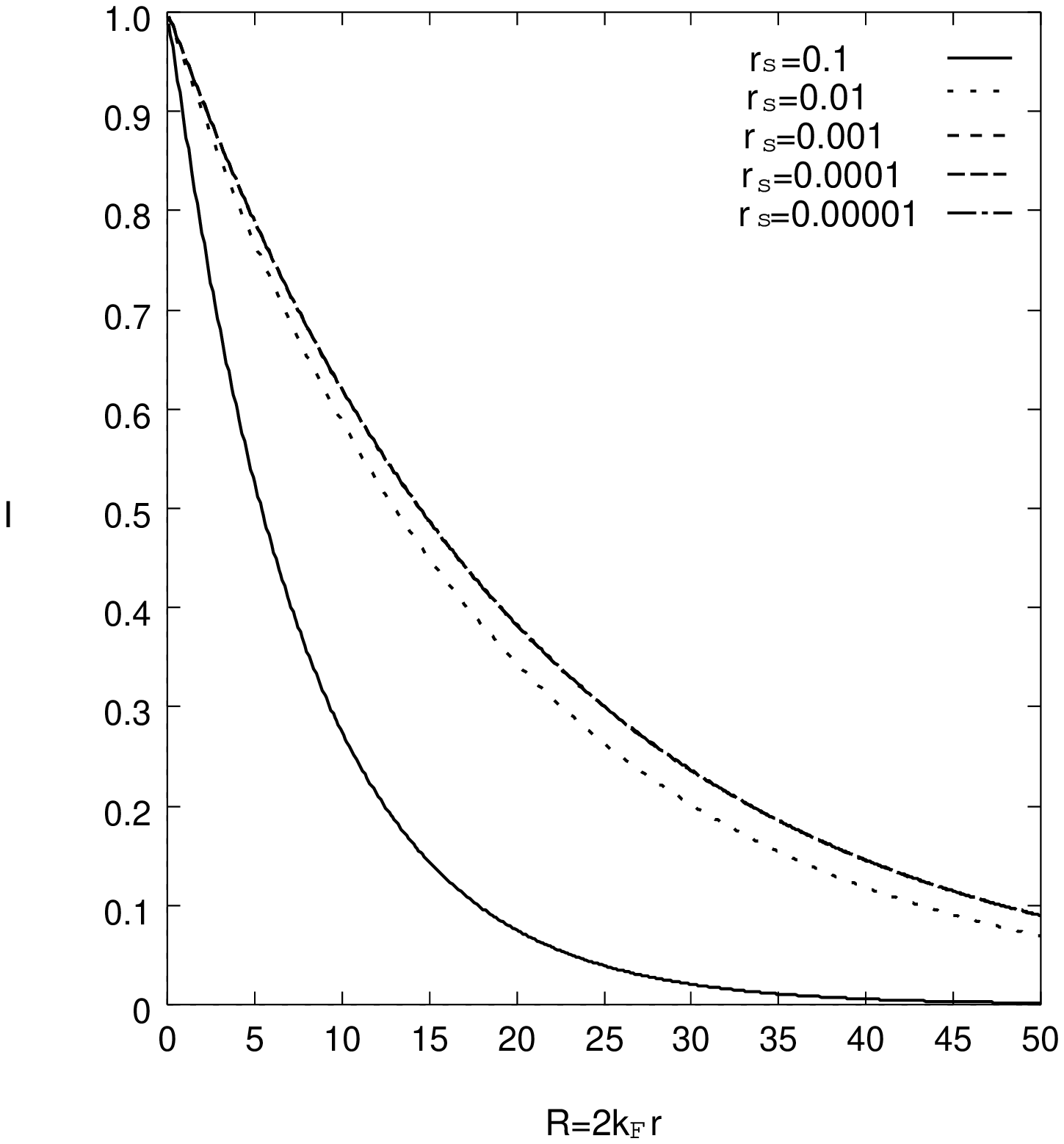}
\caption{$I$ as a function of $R=2k_{F}r$ for $r_{s}$=0.1, 0.01, 0.001,
         0.0001, 0.00001.}
\end{figure}

\clearpage
\begin{figure}
\epsscale{0.8}
\plotone{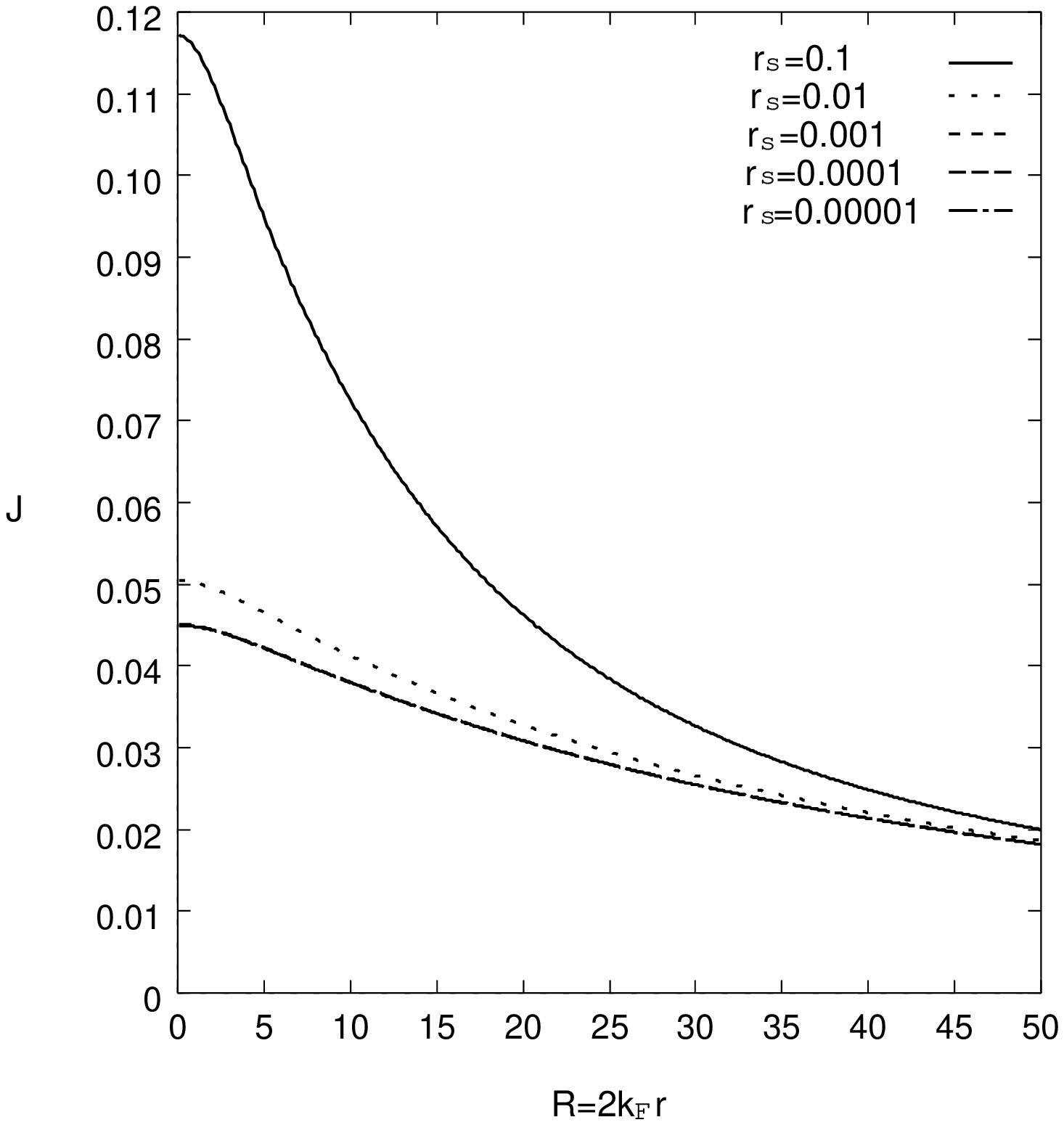}
\caption{$J$ as a function of $R=2k_{F}r$ for $r_{s}$=0.1, 0.01, 0.001,
         0.0001, 0.00001.}
\end{figure}

\clearpage
\begin{figure}
\epsscale{0.8}
\plotone{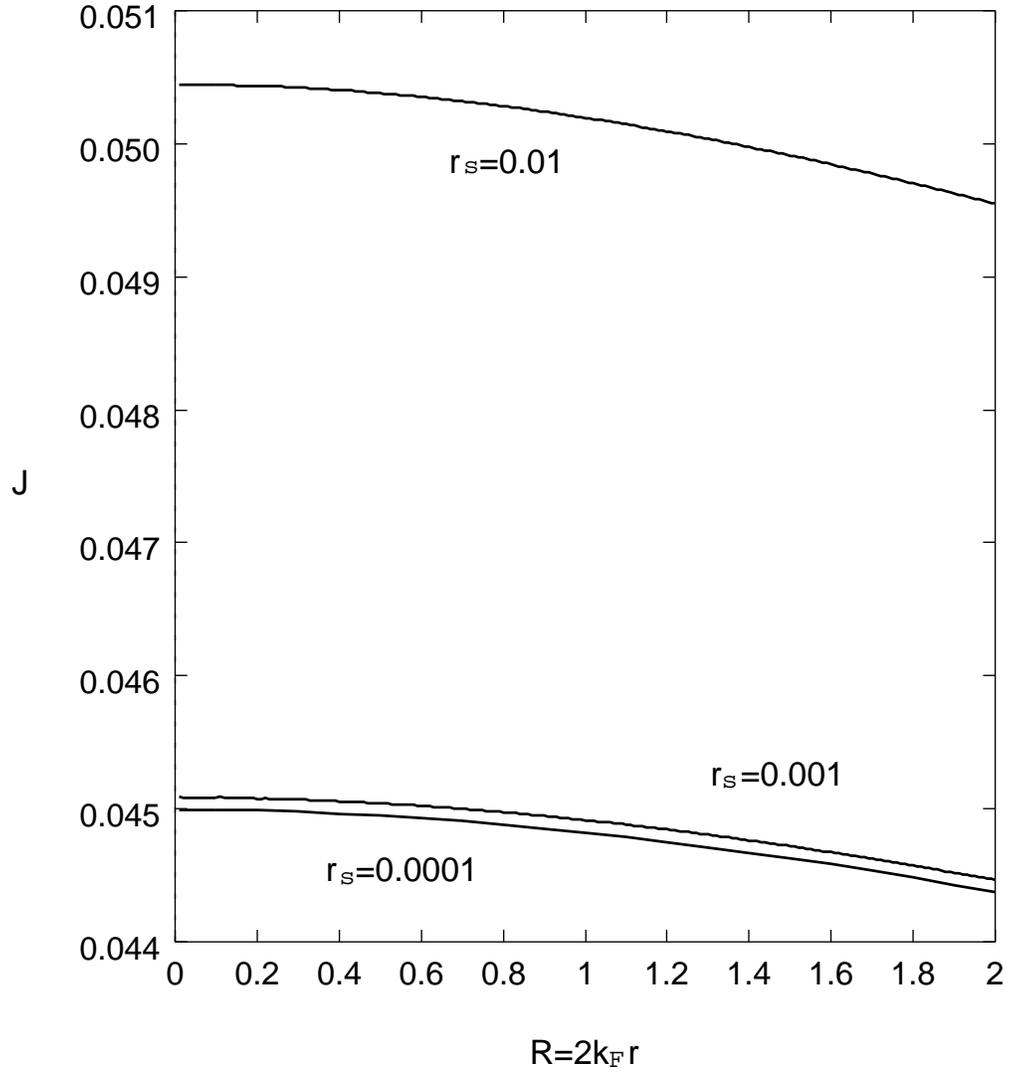}
\caption{Same as Fig.~2, but for small values of $R=2k_{F}r$.}
\end{figure}

\clearpage
\begin{figure}
\epsscale{0.8}
\plotone{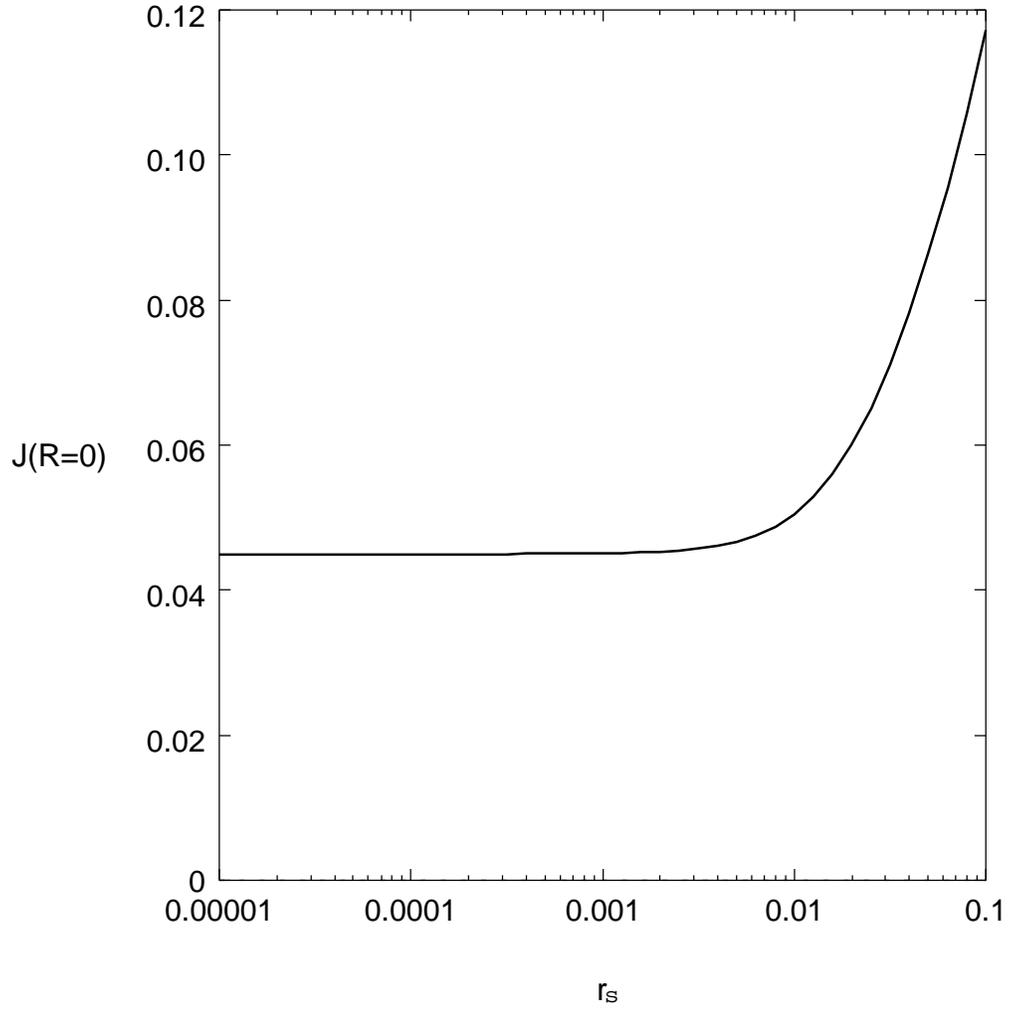}
\caption{$J(R=0)$ as a function of $r_{s}$.}
\end{figure}

\end{document}